\documentclass[11pt]{jacs}
\title{A Comparative Study  of Structural, Acidic and Hydrophobic properties 
 of Sn-BEA with Ti--BEA using Periodic Density Functional Theory}
\author{\large{Sharan Shetty},$^{\dagger\S}$ 
Dilip G. Kanhere,$^{\dagger}$ 
Annick Goursot,$^{\ddagger}$ Sourav Pal$^{\S *}$ \\   
\it{Contributions from 
the Centre for Modeling and Simulation,} \\ 
\it{Department of Physics, University of Pune, Pune 411007,} \\
\it{National Chemical Laboratory, Pune 411008 and} \\ 
\it{Ecole de Chimie, Montpellier, Cedex 5 France} \\
s.pal@ncl.res.in}
\usepackage[super,sort&compress]{natbib}
\usepackage[dvips]{graphicx}
\begin{document}

\maketitle

\begin{abstract}
Periodic density functional theory has been employed to characterize the 
differences in the structural, Lewis acidic and hydrophobic properties
of Sn--BEA and Ti--BEA. We show that the incorporation of Sn increases 
the Lewis acidity of BEA compared to the incorporation of Ti. Hence, the 
present work gives an insight into
 the role of Sn in increasing the efficiency of the oxidation
reactions. 
The results also justify that the percentage of Sn substituted in BEA 
is less than Ti. The structural analysis shows that the first coordination
shell of Sn is larger than that of Ti. However, the second coordination of
both sites remains the same. Moreover, the 
water resistant properties of these substituted zeolites are quantified.  
\end{abstract}

\section{Introduction}
Zeolite beta (BEA) has been used as one of the active 
catalysts for carrying out 
several organic reactions such as epoxidation of olefins, \cite{bea1} 
aromatic and aliphatic alkylation,\cite{bea2} acid catalyzed 
reactions,\cite{bea3} etc.
Some of the important 
reactions which can be catalyzed by BEA include, the Baeyer--Villiger 
oxidation (BVO) reaction and the Meerwein--Ponndorf--Verley reduction of 
aldehydes and Oppenauer's  oxidation of alchols (MPVO) reaction.\cite{bea_bvo}
The reasons for using BEA as an efficient catalyst are its relatively large 
pore size, its flexible  framework and high acidity.\cite{appl_bea} 
It has been well established that the 
acidity of BEA can be finely tuned by the 
incorporation of various atoms such as B, Al, Ti, Zr, Fe etc. 
\cite{t_bea}$^-$\cite{ti_bea_hydro}
These sites substituted in the BEA framework 
act as active Bronsted or Lewis acid sites depending upon their 
valence states.\cite{vansanten}
Among these atoms, 
Ti--substitution in  BEA framework has proven to be an active 
catalyst for the epoxidation of
olefins in the presence of H$_2$O$_2$.\cite{ti_bea}$^,$\cite{ti_epo}
The other Ti--zeolites, which have been 
succesfully used for the oxidation of small organic molecules, are the 
titanium silicates (TS--1, TS--2).\cite{ts-1} Several studies have been
reported to understand the differences of the activity and selectivity 
between these two zeolites.\cite{bea1}$^,$\cite{diff_ts} 
Corma et al have shown that these differences 
are due to the hydrophilic/hydrophobic nature of the Ti sites.\cite{bea1}$^,$
\cite{ti_bea_hydro} 
They showed that the Ti--sites in TS are more hydrophobic
than the Al--Ti--BEA. Hence, TS was prefered over Al--Ti--BEA when  
the solvent used in the reaction is prepared in aqueous medium. 

One of the challenges in this field is to increase the efficiency of a zeolite
by substitution with other elements. 
Such an attempt has been made recently by 
incorporating Sn in BEA. Mal and Ramaswamy succesfully synthesized the
Al--free Sn--BEA.\cite{rama_bea}
 In an interesting experimental work, Corma {\it et al} showed  
that the incorporation of Sn in the BEA framework results into a more
efficient catalyst for the BVO reaction in the presence
of H$_2$O$_2$.\cite{corma_nature}
 In their study, a new mechanism was proposed for the
oxidation of ketones. They showed that the Sn site in BEA activates the
carbonyl group of the cyclohexanone followed by the attack of H$_2$O$_2$,
unlike the Ti sites which initially activate the H$_2$O$_2$. This result
was attributed to the higher Lewis acidity of the Sn site with respect to
the Ti site. 
Hence, incorporation of Sn in BEA leads to a high selectivity towards 
the formation of lactones in the BVO reaction.\cite{corma_nature}$^,$
\cite{renz_cej} 
On this background, highly selective MPVO reactions were carried out more
efficiently with Sn--BEA than Ti--BEA.\cite{corma_mpvo} In these studies,
it was shown that the Sn site is situated within the framework and no
extraframework Sn was detected. 
Although much of the experimental studies have
focused on the efficiency of the Sn--BEA, the higher Lewis acidity of the 
Sn site compared to the Ti site in BEA is still not known. Recently, 
Sever and Root used the M(OH)$_4$ (M = Sn, Ti) cluster models to 
investigate the reaction pathways for the BVO reaction.\cite{sever}    

The activity and selectivity of the zeolite mainly depend on the nature of
the active sites, such as local coordination, interaction
with the incoming molecules, percentage of substitution of T atoms 
in the framework, etc. 
One of the important issues concerning the activity and selectivity 
of the zeolite is its hydrophobic/hydrophilic nature.\cite{corma_hydro}$^,$
\cite{weitkamp}  
It is known that if the zeolite is hydrophilic in nature, the water present in 
the solvent poisons the active sites. This hinders the kinetics
of the reaction and decreases the activity 
of the zeolite. Corma {\it et al} have bypassed this problem 
by modifying the catalyst design, which allows the use of Sn--BEA  in the
presence of aqueous media.\cite{sn_hydro} Very recently, Boronat {\it et al} 
have done theoretical calculations using a Sn(OSiH$_3$)$_3$OH cluster model 
to understand the effect of H$_2$O during the BVO reaction. Their results
show that one water molecule is permanently attached to the Sn active site.  
Interestingly, 
Fois {\it et al} have studied the interaction of water molecules with the 
Ti sites in Ti--Offretite using Car--Parrinello molecular dynamics.\cite{fois} 
They found that at higher loading of water molecules, the Ti atom expands its 
coordination  number. 

In the last decade, several experimental and theoretical studies 
have been employed 
to characterize the role of Ti sites at a microscopic level in various 
Ti-zeolite systems.\cite{bea1}$^,$\cite{ti_bea}$^,$\cite{dovesi}
It has been revealed that due to high crystallinity, low Ti content 
and large quadrupolar moment
of Ti, accurate information on the Ti sites in BEA is not possible through 
experimental techniques.\cite{notary}
 Hence, it is necessary to use theoretical 
methods to explore the local behavior for eg. structure, electronic and 
bonding properties of these sites.
Sastre and Corma have used {\it ab initio} calculations to discuss 
the role of the Ti sites in Ti--BEA and TS--1.\cite{sastre} 
The energies of the lowest
unoccupied molecular orbital (LUMO) 
of Ti--BEA and TS--1 with one Ti substituted in turn at every T site, 
were shown to be different. Furthermore,
the Ti--sites in Ti--BEA were found to be more acidic than in TS--1 and
this acidity varies among all the Ti--sites in both 
zeolites.\cite{sastre} 
This proves that not only different Ti--containing zeolites have 
different acidity, but also different T--sites within a particular 
zeolite would have varying acidity. Very recently, Bare et al have used
EXAFS technique to investigate the Sn--site in Sn--BEA.\cite{bare} 
They showed that
Sn was not randomly distributed in BEA, and takes specific crystallographic 
sites, i.e. T5/T6 sites in their nomenclature, which corresponds to T1 and T2 
in our nomenclature, following Newsam {\it et al}.\cite{newsam} 
Surprisingly, they found that this substitution
takes place through pairing of these sites, within the six membered ring i.e.
two T1 or two T2. 
However, no explanation was given
for this distribution. 
At the same time, in a theoretical work using a periodic approach based on
density functional theory (DFT) we characterized the Sn--sites 
in BEA.\cite{shetty} We showed that the T2--site 
would be the most probable site for the Sn substitution 
based on thermodynamics consideration of the largest 
cohesive energy in a dehydrated BEA zeolite.
Moreover, we found
that the substitution of 2 Sn atoms per unit cell was thermodynamically 
unfavorable. This was consistent with the earlier experimental results.
Parallel to this, Boronat {\it et al} carried out a cluster calculation 
of the Sn--BEA interaction with cyclohexanone and H$_2$O$_2$.\cite{boronat}
 They showed that the BVO reaction in Sn--BEA is due to the activation of 
cyclohexanone at the Sn site.
 
As it can be seen from the above description, 
the incorporation of Sn in BEA proves it to be a 
better catalytic site than Ti. Hence, a detailed information
on the differences in the properties of Sn and Ti sites in BEA, such as
the quantification of the Lewis acidity, number of T atoms to be substituted
in the unit cell and hydrophobicity, are of fundamental 
importance and are still to be resolved. 
The aim of the present theoretical study is to bring out the differences in 
these substituted BEA zeolites by analyzing their structural, 
electronic and hydrophobic properties. 
Moreover, it is also important
to understand why the framework Sn site activates the carbonyl group of 
cyclohexanone and not H$_2$O$_2$ in the BVO reaction, while Ti
behaves the other way. 
We investigate this issue on the basis of hard--soft acid--base (HSAB)
principle. 

\section{Methodology and Computational Details}
Several theoretical studies based on a classical as well as 
quantum potential have been proposed to study the properties of 
zeolites.\cite{sastre}$^,$\cite{jentys}$^{--}$\cite{rozanska} 
It has been a practise to adopt cluster models cut from the
zeolite crystals to study these properties. One of the obvious reasons
to use cluster model is that it is computationally cheap. 
Sauer {\it et al} have done an extensive study of zeolites using cluster 
models.\cite{deman}$^,$\cite{sauer1} 
However, the active site represented in the 
cluster model is in a different electronic environment 
than in which it would be in 
a crystal.\cite{sauer1}$^,$\cite{nicholas}$^,$\cite{rozanska} 
A periodic approach provides a more realistic 
description to study the properties 
of a crystal.\cite{nicholas}$^,$\cite{dovesi}
Although zeolite catalysts are neither crystals nor periodic solids, 
it is more convenient
to use periodic boundary conditions, when there are very few substituted
sites per unit cell.

\begin{figure}
\caption{Crystallographically defined 9 T sites of BEA. The
grey spheres represent the Si sites.}
\begin{center}
\includegraphics[width=0.5\textwidth]{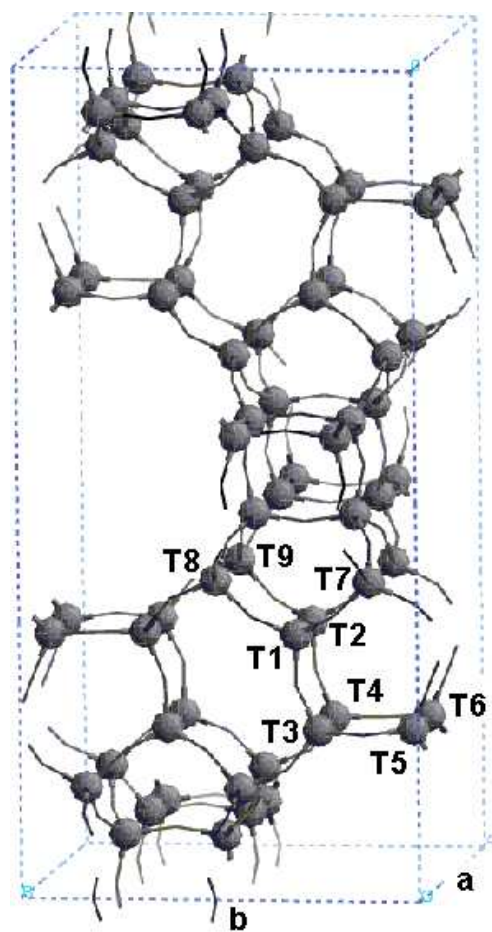}
\end{center}
\label{Fig. 1}
\end{figure}

Earlier experimental studies have indeed proved
that Sn and Ti sites in BEA are very few, that they are 
situated within the framework and during 
the BVO or MPVO reaction these sites do not dissociate from the 
framework.\cite{bea_bvo}$^{(a)}$$^,$\cite{renz_cej}
In the present work we have employed the periodic DFT to investigate the 
properties of Sn--BEA and Ti--BEA. Advantage of using periodic boundary
conditions is that the long range electrostatic interactions are included 
within Ewald summations.
The instantaneous stationary electronic ground state is 
calculated by solving the Kohn-Sham equation based on DFT. The valence
electrons have been represented by the plane waves in 
conjunction with the Vanderbilt's ultra--soft pseudopotential for 
the.\cite{vanderbilt} 
It is worth mentioning that during the interaction between two systems the
complete plane wave avoids the basis set superposition error. 
The exchange correlation functional is expressed by the 
the generalized gradient approximation (GGA) with 
the Perdew--Wang 91 functional.\cite{pw91} 
The calculations were restricted to 
the gamma point in the Brillouin zone sampling. All the calculations
have been performed by the VASP code.\cite{vasp} 

BEA is a high silica zeolite and consists of two different ordered 
polytypic series {\it viz.} polymorph A and polymorph B.\cite{newsam}
 It has two
mutually perpendicular straight channels with a cross section of 
0.76*0.64 nm which run along a and b directions. 
Intersecting to these, at right angles, 
a helical channel of 0.55 x 0.55 nm also 
exists along the c-axis. This gives rise to a three dimensional 
pore system of 12--membered ring aperture. The unit cell of an ideal fully 
siliceous BEA
consists of 192 atoms with 64 Si and 128 O atoms distributed within 
the tetragonal lattice of dimension 12.6 x 12.6 x 26.2 {\AA}. 
There are 9 distinct crystallographically defined T sites as shown in Fig. 1. 
We adopt the experimental structure as defined by Newsam {\it et al} and 
accordingly define the 9 T sites in the unit cell of BEA.\cite{newsam} 

The structural optimization of the Si, Sn and Ti--BEA have been carried out 
in two steps. In the first step, conjugate gradient method was used to 
optimize the unit cell of BEA. The optimization was considered to 
achieve when the forces on the atoms were less than 0.1 eV/{\AA}. In the 
second step, these optimized geometries were re--optimized with 
quasi--Newton method unless the forces on the atoms were less than 
0.06 eV/{\AA}. One should note that during the optimization the 
cell shape of the unit cell has been fully relaxed, while keeping its
volume constant. In the case of Sn and Ti--BEA each of the 9 distinct
T sites were substituted by Sn and Ti atoms (i.e. 
Si/(Sn or Ti) = 63/1, respectively, and were optimized. Once the active
site in Sn and Ti--BEA was confirmed, one water molecule was introduced near 
to these active sites and the same optimization procedure was followed as 
discussed above. The 
structural data for the Sn--BEA has been taken from a recent publication by 
us.\cite{shetty}   
In addition to this, we have also carried out DFT calculations of 
cyclohexanone and H$_2$O$_2$ molecules using a supercell. These systems 
were optimized by the conjugate gradient methods only, until the
forces on the atoms were less than 0.005 eV/{\AA}. 
    
\section{Results and Discussion} 
\subsection{Structure of Sn--BEA and Ti--BEA}
We have already discussed the structure and energetics of Sn--BEA in a 
recent publication.\cite{shetty} 
In the present work we briefly recall this discussion
which is necessary for comparing it with the structure and energetics of 
Ti--BEA and also to study its hydrophobic characteristic. 

Table 1 and Table 2 present the optimized 
structural details of all the 9 T sites of Sn--BEA and Ti--BEA, respectively.   
It should be noted that only the average bond distances and bond angles
are presented. It can be seen from Table 1 that the Sn--O bond distance 
range between 1.908 to 1.917 {\AA}, the Sn--O--Si bond angle range 
from 136 to 144.2 deg and the Sn--Si distance is around 3.241$\pm$0.100 {\AA}. 
Very recently, Bare {\it et al}, with the help of EXAFS
technique showed that the Sn--O bond distances and the Sn--Si distances 
in Sn--BEA were around 
1.906 {\AA} and 3.5 {\AA}, respectively.\cite{bare} 
This clearly shows that the theoretical results presented 
by us are in good agreement with the experimental results. However, the 
theoretical results of Bare {\it et al} 
were not consistent with their experimental 
data. This may be due to keeping the shape of the unit cell fixed during
the optimization and using the local density approximation 
exchange--correlation potential in their study.\cite{bare} 
On the other hand, we have relaxed the 
lattice vectors of the unit cell during the optimization and also 
used the GGA exchange--correlation potential, as explained in the 
earlier section. The change in the local coordination of the 
T site in Sn--BEA compared to the Si--BEA has been illustrated in 
the earlier study. As already mentioned above, sites
T5 and T6 of ref. 25 correspond to sites T1 and T2 in our work, in which we use
the nomenclature of Newsam et al.\cite{newsam} 

\begin{table}
\begin{center}
\caption{Optimized structural parameters of Sn--BEA. 
Average Sn--O bond lengths, Sn--O--Si bond angles and Sn--Si distances of
 all the 9 T sites of Sn--BEA.}
\begin{tabular}{|l|l|l|l|}
\hline
T--sites & Sn--O (\AA) & Sn--O--Si (deg) & Sn--Si (\AA)\\
\hline
T1  &  1.911 & 143.5 & 3.336 \\
\hline
T2  &  1.909 & 144.2 & 3.341 \\
\hline
T3  &  1.910 & 140.6 & 3.241 \\
\hline
T4  &  1.917 & 136.0 & 3.281 \\
\hline
T5  &  1.913 & 142.2 & 3.297 \\
\hline
T6  &  1.910 & 141.2 & 3.297 \\
\hline
T7  &  1.911 & 140.6 & 3.282 \\
\hline
T8  &  1.908 & 140.0 & 3.282 \\
\hline
T9  &  1.912 & 137.8 & 3.270 \\
\hline
\end{tabular}
\end{center}
\end{table}

\begin{table}
\begin{center}
\caption{Optimized structural parameters of Ti--BEA. 
Average Ti--O bond lengths, Ti--O--Si bond angles, Ti--Si distances of all the 
9 T sites of Ti--BEA.}
\begin{tabular}{|c|c|c|c|}
\hline
T--sites & Ti--O (\AA) & Ti--O--Si (deg) & Ti--Si (\AA) \\
\hline
T1 & 1.799 & 151.7 & 3.302 \\
\hline
T2 & 1.797 & 152.4 & 3.304 \\
\hline
T3 & 1.794 & 145.0 & 3.220 \\
\hline
T4 & 1.797 & 145.4 & 3.233 \\
\hline
T5 & 1.799 & 148.1 & 3.257 \\
\hline
T6 & 1.799 & 148.5 & 3.263 \\
\hline
T7 & 1.794 & 149.0 & 3.269 \\
\hline
T8 & 1.795 & 147.4 & 3.249 \\
\hline
T9 & 1.798 & 144.0 & 3.225 \\
\hline
\end{tabular}
\end{center}
\end{table}
Table 2 shows that the average Ti--O bond distances of the 9 T sites in BEA
vary from 1.794 to 1.799 {\AA}. These values are in good agreement with the
earlier works on Ti--BEA.\cite{ti_bea} 
Compared to Sn--O bond distances, the Ti--O distances
are smaller. This is due to the larger atomic size of Sn with respect to Ti. 
From the data of Tables 1 and 2, it can be noticed that the average Sn-O 
and Ti-O bond lengths are very similar for all T sites, whereas 
the corresponding bond
angles have a large range of variation. Moreover, in both Sn and Ti--BEA
models, the largest average angles belong to the T1 and T2 sites. The average
experimental values of T1--O--T and T2--O--T 
angles in the unsubstituted Si-BEA are
155.3 and 155.9 deg respectively, and they also correspond to the largest
T--O--T angles in the framework.
If we compare Sn and Ti substituted in the framework with Si, we get the
expected order for average T-O bond lengths Sn-O$>$Ti-O$>$Si-O, 
with around 0.12 to 0.15 {\AA}  difference at each replacement. 
The average T1--O--T or T2--O--T 
bond angles vary as Sn--O--Si$<$Ti--O--Si$<$Si--O--Si.

This Ti--O--Si bond angles which range between 144 to 152 deg,
are larger than the Sn--O--Si bond angles with a range between 
136.0--143.5 deg. 
Due to the angular flexibility, the Ti--Si distance differ 
only by $\sim$0.04 {\AA} from the Sn--Si
distance. Although the first coordination shell radius of Ti is
smaller than that of Sn, the second coordination shells are at similar 
distances. 
The adaptation of the BEA framework to Sn and Ti substitution 
results thus into a quite localized deformation of the siliceous framework. 
Hence, we can 
infer that the difference in adsorption properties between Sn and Ti-BEA
should be mainly due to the electronic differences of these sites.

\subsection{Energetics of Sn--BEA and Ti--BEA} 
In this subsection, we discuss the thermodynamic stability of Sn--BEA and 
Ti--BEA. This is done by calculating the cohesive energies for each of the 
9 T--sites in Sn--BEA and Ti--BEA. Cohesive energy of a solid is
defined as the difference between the energy of the bulk (solid) at 
equilibrium and the energy of the constituent atoms in their 
ground state.\cite{shetty}
Cohesive energy does not account for the kinetic formation of the system,
neither for the different nature of the synthesis intermediates
generated in aqueous solution, which can generate different routes for the
solid growth.
 
The cohesive energies of all the 9 substituted T--sites of Sn--BEA and Ti--BEA
are given in Table 3. In our earlier investigation on the energetics 
of Sn--BEA, we showed that the substitution of Sn in the BEA framework 
decreases the cohesive energy.\cite{shetty} 
Hence, the incorporation of Sn in BEA was shown to be
thermodynamically less stable than the Si--BEA. On this basis, we 
explained the fact that the incorporation of Sn in the BEA framework is 
restricted. Interestingly, Bare {\it et al} predicted the formation of
Sn pairs as the active sites, where the two Sn atoms were shown to 
be on the opposite sides of a six membered ring.\cite{bare} 
They showed that one of these pairs is present per 
8 u.c of BEA. Unfortunately, at present, it is out of 
scope to consider     
8 u.c of BEA. Nevertheless, we have carried out the calculations placing 
2 Sn atoms per u.c. at T1 and T2 (T5 and T6 according to Bare et al)
positions which are situated in the six membered ring and
are on the opposite side of each other (Fig. 1). We found that this does not 
increase the cohesive energy. 

\begin{table}
\begin{center}
\caption{Cohesive energies of all the 9 T--sites of Sn--BEA and Ti--BEA}
\begin{tabular}{|l|l|l|}
\hline
&\multicolumn{2}{l|}{Cohesive Energies (eV)}\\
\cline{2-3}
T--Sites & Sn--BEA & Ti--BEA \\
\hline
T1 & -1521.387 & -1530.797 \\ 
\hline
T2 & -1521.681 & -1530.767 \\
\hline
T3 & -1521.468 & -1530.210 \\
\hline
T4 & -1521.523 & -1530.045 \\
\hline
T5 & -1521.405 & -1530.014 \\
\hline
T6 & -1521.431 & -1530.570 \\
\hline
T7 & -1521.457 & -1530.359 \\
\hline
T8 & -1521.621 & -1530.415 \\
\hline
T9 & -1521.323 & -1530.282 \\
\hline
\end{tabular}
\end{center}
\end{table}

The cohesive energy of Si--BEA is -1527.902 eV.\cite{shetty} 
From Table 3 we see that the
cohesive energy of Ti--BEA is about 3 eV higher than that of Si--BEA. This
indicates that the incorporation of Ti in BEA is thermodynamically
more favorable than that of Sn. 
Among the 9 T--sites of Ti--BEA we found that the T1 and T2 sites
have the highest stability, and that T5 is the least stable.
We have also calculated the cohesive energy
with two Ti/u.c (i.e. Ti/Si = 2/62 per u.c). The two Ti atoms
were incorporated at two different T2 positions at a distance of 9 {\AA}. 
This showed an increase in the cohesive energy of about 3 eV compared to
one Ti/u.c. This reveals that more Ti could be incorporated in BEA than   
Sn.
 We want to stress that these calculations are carried out on a dehydrated
solid resulting from a thermodynamically driven synthesis, ignoring the
effects of the various ingredients and formation conditions, i.e. the nature
and energies of the synthesis intermediates.
Nevertheless, 
these results are consistent with the earlier experimental works, where it
has been shown that the amount of incorporated Ti is larger than that of Sn in 
BEA.\cite{ti_bea}$^,$\cite{renz_cej} 
  
\subsection{Lewis acidity of Sn--BEA and Ti--BEA}
Earlier experimental studies have conjectured that Sn acts as a better Lewis 
acidic site than Ti in BEA.\cite{corma_nature}$^{--}$\cite{corma_mpvo} 
Hence, Sn--BEA acts as a more active catalyst for the 
oxidation reactions. This motivated us to compare the Lewis acidity of
Sn and Ti--BEA. 
First, one must recall that the Lewis acidity, being related 
with an electron acceptor character, can be correlated with the global
electron affinity of the solid. Qualitatively, LUMO energies can be used for a
comparison between the electron affinities of Sn and Ti-BEA.\cite{sastre} 
The HOMO and the LUMO energies, and their corresponding HOMO--LUMO gaps of 
Sn--BEA and Ti--BEA have been reported in Table 4. 
Globally, the average LUMO energy among the Sn substituted models is lower
than that for the Ti ones.
In our earlier 
results on Sn--BEA, we have shown that out of the 9 T--sites the 
T1 and the T2 sites have low LUMO energies compared to the other T--sites, 
and would be the 
probable sites for the reaction.\cite{shetty} 
Interestingly, T1 and T2 have been 
proposed as the most probable sites for Sn substitution from EXAFS 
experiments.\cite{bare}
The two corresponding LUMO's have similar low energies, making
these two models good candidates as Lewis acids. Both sites have also the
smallest HOMO-LUMO gap. A smaller gap, in a solid, correlates with a larger
global softness. The most probable Sn-BEA solids would thus correspond to the
most Lewis acidic and the more ``soft" models.

\begin{table}
\begin{center}
\caption{Energies of the HOMO, LUMO and the HOMO--LUMO gaps 
of the 9 T--sites of Sn--BEA and Ti--BEA} 
\begin{tabular}{|l|l|l|l|l|l|l|}
\hline
\multicolumn{1}{|c|}{T-Sites}&\multicolumn{3}{c|}{Sn--BEA}&\multicolumn{3}{c|}{Ti--BEA} \\
\cline{2-7}
& HOMO (eV) & LUMO (eV) & Gap (eV) & HOMO (eV) & LUMO (eV) & Gap (eV) \\
\hline
T1 & -3.124 & 1.333 & 4.457 & -3.135 & 1.417 & 4.552 \\
\hline
T2 & -3.125 & 1.366 & 4.491 & -3.133 & 1.469 & 4.602 \\
\hline
T3 & -3.131 & 1.557 & 4.688 & -3.121 & 1.548 & 4.669 \\
\hline
T4 & -3.117 & 1.421 & 4.538 & -3.120 & 1.492 & 4.612 \\
\hline
T5 & -3.131 & 1.450 & 4.581 & -3.152 & 1.500 & 4.652 \\
\hline
T6 & -3.120 & 1.426 & 4.546 & -3.145 & 1.453 & 4.598 \\
\hline
T7 & -3.121 & 1.419 & 4.540 & -3.156 & 1.486 & 4.642 \\
\hline
T8 & -3.117 & 1.497 & 4.614 & -3.144 & 1.470 & 4.620 \\
\hline
T9 & -3.114 & 1.506 & 4.620 & -3.121 & 1.454 & 4.575 \\
\hline
\end{tabular}
\end{center}
\end{table}

In the case of Ti--BEA, we can see from Table 4 that the T1 site has the 
lowest LUMO energy, whereas T3  has the highest. We can also notice that 
T1 and T2, which have the largest cohesive energies, have also low HOMO-LUMO
 gaps, T1 having the smallest.
Considering these two factors together, we propose that these sites would
be also the most favorable sites for the substitution by Ti and  
also for the reaction to take place. 
We propose thus that, in both cases, Sn and Ti would be more probably
substituted at the T1 and T2 sites. Considering their LUMO energies, about 0.1
eV lower for Sn-BEA, we can infer that Sn--BEA is more Lewis acidic 
than Ti--BEA. Moreover, the corresponding HOMO-LUMO gaps being lower for Sn-
BEA than for Ti-BEA, this also suggests that Sn-BEA is a softer acid. This
conclusion is also supported by the following trend: whereas the cohesive
energies of the T1-T9 substituted Sn-BEA solids spreads on 0.36 eV, those of
the Ti-BEA solids spread over 0.66 eV. Despite its smaller radius, Ti has thus
less ability to adapt to the various geometric environments, showing the
behaviour of a "harder" species.

\subsection{Hydrophobicity of Sn--BEA and Ti--BEA}  
One of the important issues concerning the selectivity towards the
organic molecules in zeolites, is the hydrophobic character of these 
catalysts.\cite{corma_hydro}
Indeed, for reactions such as BVO and MPVO in the
presence of aqueous solvents, 
zeolites containing both, Lewis acidity and hydrophobicity would
be the most appropriate.\cite{sn_hydro}$^,$\cite{boronat} 
In fact, being a product of reaction, water is always present in the catalyst
pores. However, this presence is not desirable, because its adsorption is
competitive with that of reactants and also due to the product hydrolysis. On
a perfect silicate surface, water is physisorbed, i.e., its interaction energy
is weak, mainly due to van der Waals forces. As soon as defects are present,
water may bind to the silanols or dissociate and react with the surface
\cite{ma}
In order to be
hydrophobic, zeolites must thus present less or no defects. 
If this is achieved, i.e.
for highly hydrophobic samples, experimental results show that substituted Ti-
BEA is much more hydrophobic than Sn-BEA.\cite{sn_hydro}
Although it is hardly possible to compare Ti-BEA and Sn-BEA with a high
loading of water, it is
of particular interest to investigate, at the microscopic level,
 the coordination of
Sn and Ti sites in presence and absence of one water molecule.
For this comparison, Sn and Ti have been located at
sites T2 and T1, respectively.
The full systems have then been optimized.

Table 5 gives the averaged optimized T--O(BEA), T--OH$_2$ 
bond lenghths, T--O--Si
bond angles and the T--Si distances, where T = Sn and Ti. We can see that
after hydration, the Sn--O distance has been increased by 0.014 {\AA} and the
Sn--O--Si angle is also increased by about 2.3 degs 
with respect to the dehydrated Sn--BEA.
The bond distance between the Sn site and the H$_2$O
is 2.41 {\AA}.  
The hydrated Ti--BEA shows a similar trend with a
Ti--O bond length and the Ti--O--Si bond angle which have been increased 
by 0.019 {\AA} and 2.9 degs, respectively. The Ti--OH$_2$ bond distance is 
2.35 {\AA}. We see that the Sn--OH$_2$ distance is longer than Ti--OH$_2$.
In order to understand the adsorption of the H$_2$O molecule to the T sites,
we have calculated the binding energy (B.E.)(Table 5). This
is done as follows

\begin{displaymath}
B.E. = E_{complex}(BEA+H_{2}O)-\{E(BEA)+E(H_{2}O)\}         
\end{displaymath}

As can be seen from Table 5, the B.E. is positive for both systems.
This shows that the  
formation of the complex is less stable
than the separate entities (endothermic), and that water
molecules do not like to form a stable complex with either of the sites
{\it viz.} Sn and Ti in BEA.
Furthermore, the Ti--BEA and H$_2$O complex is $\sim$3 kJ/mol 
less stable than the Sn--BEA and H$_2$O complex. 

\begin{table}
\begin{center}
\caption{Structural parameters and the 
binding energies (B.E.) of Sn--BEA and Ti--BEA in the presence of 
H$_2$O}
\begin{tabular}{|l|l|l|}
\hline
& Sn--BEA + H$_2$O & Ti--BEA + H$_2$O \\
\hline
T--O(BEA) (\AA)  & 1.923 & 1.818 \\
\hline
T--O--Si (deg) & 146.50 & 154.66 \\
\hline
T--Si (\AA) & 3.369 & 3.330 \\
\hline
T--OH$_2$ (\AA) & 2.412 & 2.350 \\
\hline
B. E. (kJ/mol) & 4.87 & 7.82 \\
\hline
\end{tabular}
\end{center}
\end{table}

These results are in qualitative agreement with the experimental findings that
Sn- and Ti-BEA are hydrophobic Lewis acid catalysts. However, they may also
appear surprising since adsorption of one water molecule has been reported on
Ti-zeolites with low but exothermic interaction energies.\cite{fois}
 This difference may be
due to the nature of the zeolite framework (beta versus offretite) or to the
type of calculations (cluster versus periodic).
It must be recalled that interaction energies calculated with DFT based
methods do not include van der Waals attractive contributions. In recent work,
these dispersion terms have been added empirically\cite{wu} 
or using
adequate correlation functionals.\cite{dion}
It is easy to
give an a posteriori estimate of the van der Waals stabilization of water
bound to the Sn or Ti sites in BEA, using an empirical correction. Using our
optimized Sn and Ti structures, the van der Waals stabilization energy of the
bound water molecule has been calculated using the universal force 
field.\cite{rappe} 
The following energies have been found: -3.3kcal/mol for Sn-BEA
and -2.4 kcal/mol for Ti-BEA. Since these empirical van der Waals terms are
additive, one can infer that a water dimer would form a very low exothermic
complex with the Sn-BEA model, but would still be non bonding with the Ti
model.
Hence, these results show that 
Ti--BEA is more hydrophobic than Sn--BEA. This confirms the earlier 
experimental findings.\cite{sn_hydro}$^{(a)}$   
 
\subsection{Reactivity of Sn--BEA and Ti--BEA towards cyclohexanone 
and H$_2$O$_2$}
We have applied the HSAB principle to understand the reactivity of Sn-- and 
Ti--BEA towards cyclohexanone and H$_2$O$_2$. Pearson formulated the 
concept of HSAB principle for understanding the reactivity of 
chemical systems, and their interactions.\cite{pearson} 
This gave a new insight in 
interpreting the reactivity of chemical systems on the 
basis of their HOMO and LUMO energies.\cite{klopman} 
The systems can be categorized as 
soft acid (SA) with low lying LUMO, soft base (SB) with high lying HOMO,
hard acid (HA) with high lying LUMO and hard base (HB) with low lying HOMO.
 It has been well established that the interactions between SA--SB are 
covalent, HA--HB are ionic and SA--HB or HA--SB are mostly weak electrostatic
 and form Lewis adducts. 

\begin{table}
\begin{center}
\caption{HOMO and LUMO energies of cyclohexanone and H$_2$O$_2$}
\begin{tabular}{|l|l|l|}
\hline
& HOMO & LUMO \\
\hline
Cyclohexanone & -5.066 & -1.333 \\
\hline
H$_2$O$_2$ & -5.730 & -1.546 \\
\hline
\end{tabular}
\end{center}
\end{table}

In earlier experimental studies it has been proposed that the Sn site in BEA
polarizes the carbonyl oxygen atom of cyclohexanone 
and forms a Lewis adduct.\cite{corma_nature}$^,$\cite{renz_cej}
From above, we 
have shown that Sn-BEA behaves as a SA, compared with Ti-BEA.
Table 6
presents the HOMO and LUMO energies of cyclohexanone and H$_2$O$_2$. We
observe that the HOMO energy of cyclohexanone is $\sim$60 kJ/mol 
lower in energy than the HOMO of H$_2$O$_2$. Hence, cyclohexanone and 
H$_2$O$_2$ are HB and SB, respectively. From the HSAB prnciple we infer that 
the Sn--BEA interacts with the cyclohexanone molecule to form a SA--HB 
complex or a Lewis adduct.      

\section{Conclusions} 
The present theoretical investigation reveals the differences between the 
Sn--BEA and Ti--BEA based on their structural, Lewis acidic and hydrophobic 
properties. Our analysis shows that the Sn and Ti atoms may occupy T2 and/or
T1 crystallographic positions in BEA. Although the first  
coordination shell of Sn is larger than Ti, the second coordination shell
in both model zeolites is similar. This explains the relaxation of the local 
environment of the substituted site.
The structural data on Sn--BEA and Ti--BEA
presented in this work are in good agreement with the earlier experimental 
studies. The cohesive energy results demonstrate that the incorporation
of Ti is more favorable than Sn in BEA. Nevertheless, we 
show that Sn--BEA is more Lewis acidic than Ti, and hence proves to be a
more efficient catalyst for the oxidation reactions than Ti--BEA. One of 
the important aspects concerning the activity and selectivty of the zeolite
which we have addressed in the present work, is the water resistant
property of the Sn--BEA and Ti--BEA. A stable 
interaction of H$_2$O with 
the active sites of Sn--BEA and Ti--BEA is more favorable with Sn--BEA. Hence,
the hydrophobic property of the Sn--BEA and Ti--BEA
zeolites is predicted, as well as their comparison. 
This clearly justifies the water resistant Lewis acidic sites in 
Sn--BEA. We also extend our analysis to explain that the interaction of the
water molecule with the Ti and the Sn sites in BEA is a mere 
physisorption rather than chemisorption. 
Furthermore, we use the HSAB principle to interpret the 
formation of Lewis adduct between the Sn site and the cyclohexanone.

The present work gives thus an insight into the microscopic properties
of the active sites in Sn--BEA and Ti--BEA and the differences between them, 
which would have been otherwise difficult to  
understand through experimental methods.

\section{Acknowledgements}
We partially acknowledge the Indo-French Centre for Promotion of Advanced Research (IFCPAR) for providing the computational facilities. \\ 
$^{\dagger}$University of Pune.\\$^{\ddagger}$Ecole de Chimie.\\
$^{\S}$National Chemical Laboratory.\\
$^*$Corresponding Author

\pagebreak

\begin{figure}[p]
\begin{center}
Table of Content\\
Sharan Shetty, Dilip G. Kanhere, Annick Goursot, Sourav Pal 
\end{center}
\begin{center}
\includegraphics[width=0.5\textwidth]{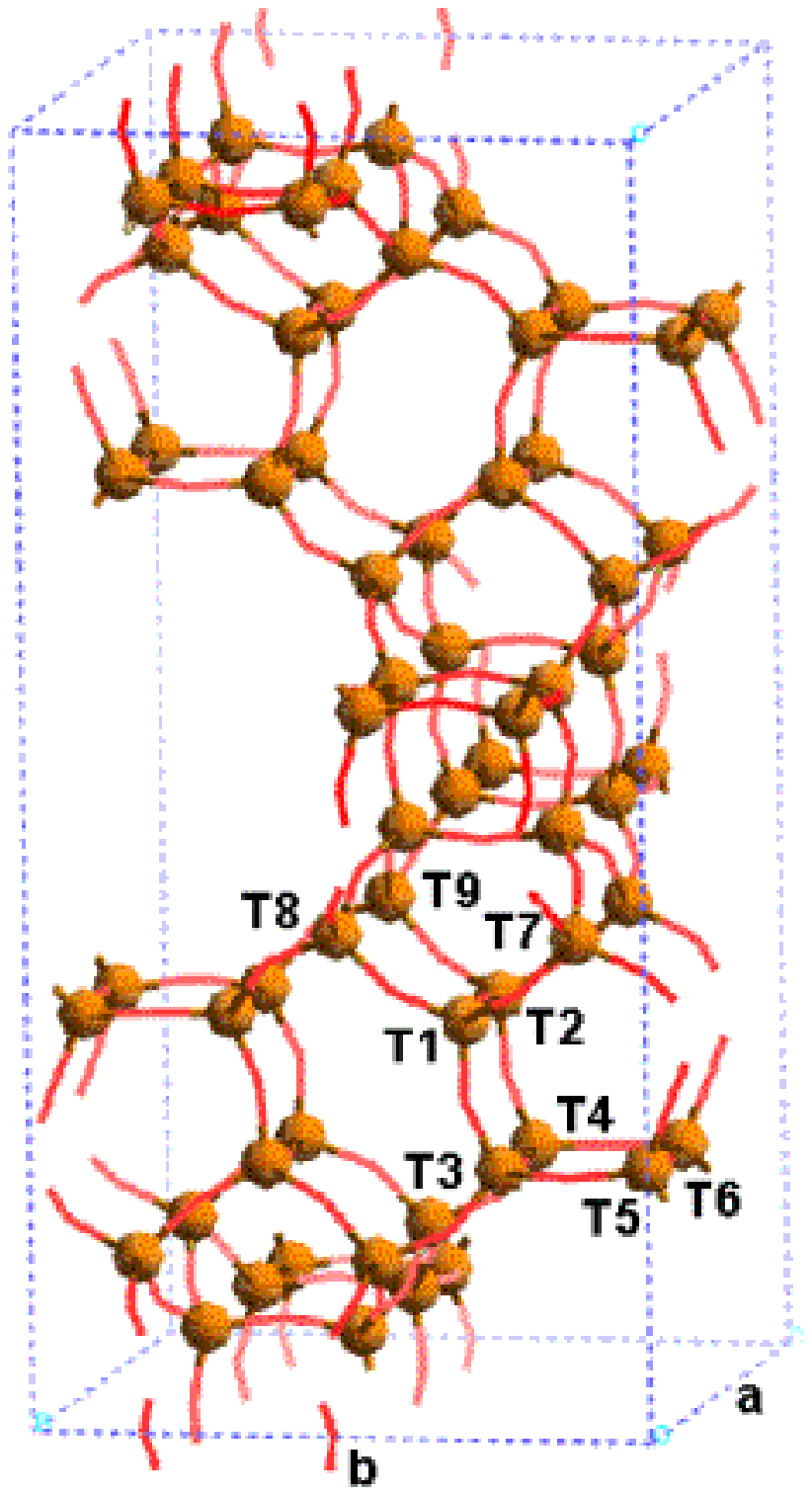}
\end{center}
\end{figure}
\end{document}